\documentclass[doublecol]{epl2} 

\usepackage{graphicx}
\usepackage{graphics}
\usepackage{epsfig}
\usepackage{color}
\usepackage{amssymb}
\usepackage{amsmath}
\usepackage{bm}
\usepackage{wasysym}
\usepackage{pifont}

\title{Lower bound on the four-point dynamical susceptibility:\\ Direct experimental test on a granular packing}
\shorttitle{} 

\author{F. Lechenault\inst{1} \and O. Dauchot\inst{1} \and G. Biroli\inst{2} \and J.~P. Bouchaud\inst{3} }
\shortauthor{F. Lechenault \etal}

\institute{                    
  \inst{1} CEA Saclay/SPEC, URA2464, L'Orme des Merisiers, 91 191 Gif-sur-Yvette, France\\
  \inst{2} CEA Saclay/SPhT, UMR2306, L'Orme des Merisiers, 91 191 Gif-sur-Yvette, France\\
  \inst{3} Science \& Finance, Capital Fund Management, 6-8 Bd Haussmann, 75009 Paris, France
}

\abstract{
We track the motion of a horizontally vibrated amorphous assembly of bidisperse hard disks, for densities ranging across the jamming transition. We derive on very general grounds a bound on the dynamical susceptibility in terms of the response of the dynamics to a change in density. This generalizes a similar bound recently derived for equilibrium liquids. 
We find that in our experimental system the bound is tight and reproduces the non-monotonic behavior of the dynamical susceptibility both in time and density across the jamming transition. The underlying scaling behavior reveals an intimate connection between anomalous diffusion and dynamical heterogeneity.}

\begin{document}

\maketitle

\section{Introduction}
The sudden change from a homogenous liquid to an amorphous rigid state is a phenomenon common to many different systems, among which supercooled liquids, colloids and granular media. This transition, called in these different contexts either a glass or a jamming transition, has very peculiar properties~\cite{DeBenedettiStillinger,Birolinewsandviews,OHern}. In particular, the associated dramatic increase of relaxation times in a very narrow window of temperatures or densities is apparently not related to the growth of any static long range order: simple static correlation functions remain quite featureless.  On the other hand, growing long range order has been discovered in {\it dynamical} correlations \cite{ediger,andersen}. A complete characterization of these so-called ``dynamical heterogenities'' has become a central  task in the context of glass and jamming transitions. Indeed, many theories ascribe the slowing down of the dynamics to some kind of growing dynamical  correlations or cooperativity, starting from the seminal work of Adam and Gibbs \cite{AG}.  Precise characterization of the amplitude of these heterogeneities  has been obtained using the recently introduced dynamical susceptibility $\chi_4$~\cite{franz,parisi}, mostly in  numerical work~\cite{andersen} but only in very few experimental situations~\cite{dauchot,Abate}. Since  evaluating this susceptibility requires extensive statistics on time-resolved density correlators, it turns out to be out of reach in most molecular glassy systems.  Instead, it has been argued that the response of the dynamics to a change in a relevant control parameter (density, temperature,...) provides  the dynamical susceptibility with a  lower bound~\cite{biroli,berbiboula,berbibouko} which potential interest relies on the fact that it only involves readily measurable two-point functions. However, the quality of this lower bound has never been  experimentally investigated and remains to be established. Only numerical results on model glass formers are available~\cite{berbibouko}.

In this letter, we focus on this issue in the general context of driven athermal systems displaying a jamming transition. First, we show that the lower bound to the dynamical susceptibility obtained for equilibrium systems in~\cite{biroliberbiboula} can be extended to driven athermal situations under a very general hypothesis. Second, we provide the first {\it experimental} test for the quality of this bound in a glassy granular monolayer undergoing a jamming transition. We directly measure the four-point dynamical susceptibility and compare it to its lower bound. We find that this bound is tight and reproduces the non-monotonic behavior of $\chi_4$ both in time and density across the transition.

\section{Theoretical context}
Consider a large bulk region of size $V$ of a still much larger stationary system. The system can be either in thermal equilibrium or a driven athermal system such as the vibrated grain assembly studied below. We assume that the density  $\rho=N/V$ is the only conserved quantity, where $N$ is the number of grains in the volume $V$. This is indeed the case in the experimental situation we focus on. 
Because of the local conservation law, $\rho$ evolves on slow time scales, through boundary effects. On very long time scales (or on different regions of size $V$), however, the density fluctuates around its mean value $\langle \rho \rangle$, with fluctuations of order $1/\sqrt{V}$. Take now any one-time {\it intensive} observable $O(t)$ and consider the joint distribution of $O(t)$ and $\rho(t)$, $P(O,\rho)=P(O|\rho) \times P(\rho)$. In absence of long-range correlations, one expects on general grounds that for large $V$ (i) the probability laws reduce to large deviation functional and (ii) the probability law for $O$ in a bulk region far from the boundaries depend on an external control parameter $\alpha$ only through the density.[$\alpha$ denotes e.g. the position of a piston, that allows one to change the average density of the system.] This can be indeed verified for equilibrium systems and for some simple model of athermal systems (see~\cite{bedaudro} for a detailed discussion). These general assumptions translate into $\log P(O|\rho)= V f(O,\rho)$ and $\log P(\rho)= V g(\rho, \alpha)$ where $f$ and $g$ are large deviation functional whose detailed expression is not known in general, but are expected (generically) to be regular and have locally quadratic maxima. We claim that these general results also hold for {\it two-point} correlation functions $C(t,t+\tau)$, provided $\tau$ is so small that $\rho$ can be considered constant within the interval $[t,t+\tau]$, such that the dynamics is slaved to a single value of $\rho$ during this time interval. In this case, $C(t,t+\tau)$ can be treated as a single time observable and we also have $\log P(C|\rho)= V f(C,\rho)$.

The dynamical susceptibility introduced in the context of supercooled liquids corresponds to the fluctuations $\delta C = C - \langle C \rangle$: $\chi_4=N\langle \delta C^2 \rangle$ where $N=\langle \rho \rangle V $ and $\left\langle \bullet\right\rangle$ stands for ensemble average. It can be easily computed from the large deviation functional by expanding around the most probable values, leading to Gaussian integrals. Following the discussion in ~\cite{berbibouko}, one obtains the following general relation:
\begin{equation}\label{eqa}
\chi_4=N\left. \langle \delta C^2 \rangle \right|_{\rho}+ N \langle \delta\rho^2 \rangle \left( \frac{\partial \langle C(t,0)\rangle}{\partial \langle \rho \rangle }\right)^2
\end{equation}
This exact equation expresses the dynamical susceptibility as a sum of fluctuations in a restricted ensemble where the density is strictly 
fixed plus a term corresponding to the dynamical fluctuations induced by density fluctuations. Dropping the first term, which is clearly positive, one finds the inequality: 
\begin{equation}
\chi_4\ge \frac{N \langle \delta\rho^2 \rangle}{\langle \rho \rangle ^2} \left( \frac{\partial \langle C(t,0)\rangle}{\partial \log \langle \rho \rangle }\right)^2.
\label{ine}
\end{equation}
In the following we shall measure directly {\it all} the quantities that appear in Eq.~(\ref{ine}) and evaluate both the RHS and the LHS of the inequality in a horizontally vibrated amorphous assembly of bidisperse hard disks in order to examine whether the RHS provides not only a bound but a quantitative estimation of the LHS. 

\section{Experimental results}
For the purpose of clarity, we first recall the essential features of our experimental system, which have been already described in~\cite{ledaubibou}. A 1:1 bidisperse monolayer of 8500 brass cylinders of diameters $d_{small} = 4\pm0.01$ mm and $d_{big} = 5\pm0.01$ mm laid out on a horizontal glass plate. The plate is subjected to in-plane oscillation at a frequency of 10 Hz, with a peak-to-peak amplitude of 10 mm. The grains are confined in a cell that is fixed in the laboratory frame. The cell has width $L \approx 100$ $d_{small}$, and its length can be adjusted by a lateral mobile wall controlled by a $\mu$m accuracy translation platen, which allows us to vary the packing fraction of the grains assembly by tiny amounts ($\delta \phi/\phi \sim 5 \, 10^{-4})$. The stroboscopic motion of a set of 1500 grains in the center of the sample is tracked by a CCD camera synchronized with the plate. The protocol, detailed in~\cite{ledaubibou}, leads to reversible stationary states for all studied packing fractions, which range from $\phi=0.8402$ to $\phi=0.8454$. For each packing fraction, the grains trajectories $\vec r_i(t)$ are recorded for twenty minutes. The averages are performed over the corresponding $10000$ configurations of the packing. Lengths are measured in $d_{small}$ units and time in cycle units. 

\begin{figure}[h!]
\centering
\includegraphics[width=\columnwidth]{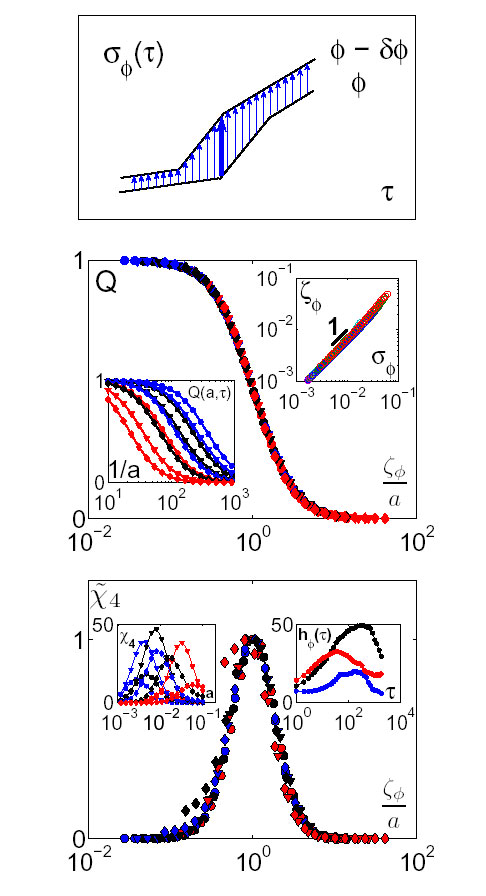}
\caption{{\bf Top:} Sketch for the non-monotonic behavior of $\partial\zeta^2$. The diffusion length $\sigma_{\phi}$, proportional to $\zeta_{\phi}$, overshoots at intermediate times, giving rise to a temporal maximum in its density derivative. {\bf Middle:} $Q\left(a,\tau\right)$ as a function of scaled variable $\zeta_{\phi}\left(\tau\right)/a$ for $\tau=12\left(\Circle\right),108\left(\triangledown\right),1047\left(\Diamond\right)$ and $\phi=0.8402\left(\text{red}\right),0.8417\left(\text{black}\right),0.8426\left(\text{blue}\right)$. Lower insert: rough functions; upper insert: characteristic length scale $\zeta_{\phi}$ vs. $\sigma_{\phi}$ for all lags and densities. The dependence is linear and the slope $\zeta_{\phi}= 1.62\sigma_{\phi}$ does not depend on $\phi$. {\bf Bottom:} Shape function $\tilde{\chi}_4$ as a function of scaled variable $\zeta_{\phi}\left(\tau\right)/a$ for the same set of data. Lower insert: $\chi_4\left(a,\tau\right)$ as a function of $a$ for the three lags above; upper insert: the amplitude $h_{\phi}\left(\tau\right)$ as a function of lag $\tau$ for the three densities above.}
\label{figone}
\vspace{0mm}
\end{figure}

At the jamming density, $\phi_J\simeq 0.842$, where mechanical rigidity sets in, long range correlations develop both in time and space and give rise to a strong peak in the dynamical susceptibility~\cite{ledaubibou}. Noticeably, the individual displacements themselves are much smaller than the grain size. A sketch of the behavior of the diffusion length $\sigma_{\phi}\left(\tau\right)\equiv\left\langle\left\|\vec r_{i}(t+\tau)-\vec r_{i}(t)\right\|^2\right\rangle_{i,t}$ is displayed in Fig. \ref{figone} (for a detailed discussion see \cite{ledaubibou}). Remarkably, the early subdiffusive regime connects to the late diffusive one through a transient super-diffusive regime. This regime, observed at characteristic time $\tau_{sD}$, emerges when approaching the jamming transition. A consequence of this striking behavior is that the variation of the diffusion length with the packing fraction, $\delta \sigma(\tau)$ is larger at $\tau_sD$ leading to a non monotonic variation of $\delta \sigma(\tau)$ as a function of time. The similarity with the behavior of $\chi_4$ is not accidental. We shall show in the following that the two turn out to be strictly related.  

We now concentrate on the evaluation of both terms of the inequality~(\ref{ine}). For that purpose, we shall focus on a particular correlation function defined as:
\begin{equation}
C\left(a,\tau\right)\equiv\frac{1}{N}\sum_i e^{-\frac{\left\|\vec r_{i}(\tau)-\vec r_{i}(0)\right\|^2}{2a^2}}
\label{defq}
\end{equation}
\noindent
where $a$ is the length scale over which the motion is probed. The tagged particle correlator $Q\left(a,\tau\right)=\langle C\left(a,\tau\right) \rangle$, when represented as a function of the probe $a$ for a given $\tau$, is characterized by a decay \textsl{length} $\zeta_{\phi}\left(\tau\right)$. This length scale is found to be proportional to the diffusion length $\sigma_{\phi}\left(\tau\right)$, with the same constant of proportionality for all the investigated densities (Fig.~(\ref{figone})-Middle). The correlator $Q$ hence follows the approximate scaling:
\begin{equation}
Q\left(a,\tau\right)=\tilde{Q}(x) \quad {\mbox{with}} \quad x = \frac{a}{\zeta_{\phi}(\tau)},
\label{scq}
\end{equation}
\noindent
which accounts for a corresponding scaling of the displacements distribution probabilities. Perhaps surprisingly, the dynamical susceptibility $\chi_4=N\langle \delta C^2 \rangle$ partly inherits this scaling property and reads
\begin{equation}
\chi_4\left(a,\tau\right)=h_{\phi}\left(\tau\right)\tilde{\chi}_4(x),
\label{scx}
\end{equation}
\noindent 
where we normalize the maximum value of $\tilde{\chi}_4$, reached for $x=1$, to be equal to $1$ (see fig.~\ref{figone}-Bottom). The shape function $\tilde{\chi}_4$ fully encodes the dependence on the probe lengthscale, whereas the amplitude $h_{\phi}(\tau)$ encodes the lag dependence of the heterogeneity of the dynamics for each packing fraction and reaches a maximum near $\tau_{sD}$. 

\begin{figure}[t]
\centering
\includegraphics[width=\columnwidth]{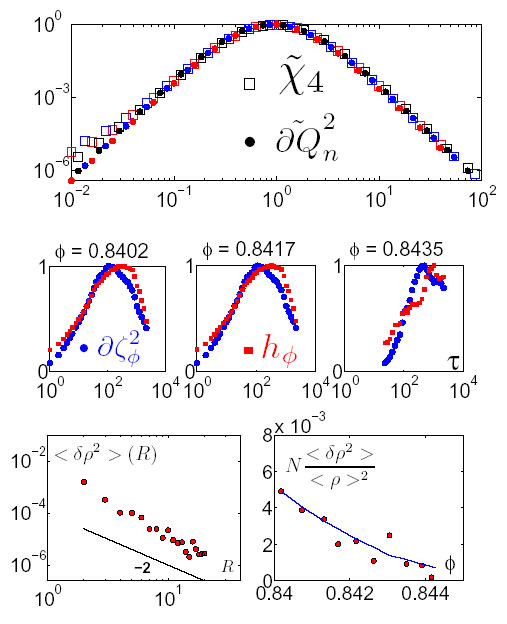}
\caption{{\bf Top:} Shape functions $\tilde{\chi}_4\left(\square\right)$ and $\tilde{\partial Q}^2_n\left(\newmoon\right)$ as a function of scaled variable $x$ for $\phi=0.8402\left(\text{red}\right),0.8417\left(\text{black}\right),0.8426\left(\text{blue}\right)$. The index $n$ signals that the function has been normalized by its maximum for comparison. {\bf Middle:} Amplitudes $h_{\phi}\left(\blacksquare\right)$ and $\partial\zeta_{\phi}\left(\newmoon\right)$  as a function of lag $\tau$ for three packing fractions $\phi=0.8402,0.8417,0.8435$. {\bf Bottom left:} density fluctuation $\langle \delta \rho ^2 \rangle \left(R\right)$ as a function of the radius $R$ of the disk within which it is evaluated, for $\phi=0.8417$. The black line corresponds to the expected $R^{-2}$ behavior. {\bf Bottom Right:} Normalized density fluctuations ${N\langle \delta \rho^2 \rangle}/{\left\langle\rho\right\rangle^2}$ as a function of packing fraction. The blue line is the interpolation we used to evaluate the bound.}  
\label{Figshape}
\vspace{0mm}
\end{figure}

Plugging the scaling properties (\ref{scq},\ref{scx}) into inequality (\ref{ine}) and using the chain rule yields
\begin{equation}
h_{\phi}\left(\tau\right)\tilde{\chi}_4\left(x\right)\geq\frac{N\langle \delta \rho^2 \rangle}{\left\langle\rho\right\rangle^2}\left(\frac{\partial\log{\zeta_{\phi}\left(\tau\right)}}{\partial\log{\rho}}\right)^2\left(\frac{\partial \tilde{Q}\left(x\right)}{\partial \log{x}}\right)^2.
\end{equation}
\noindent
In this expression, one obtains a clear factorization of the probe dependence, contained in the shape functions, respectively:
\begin{equation}
\tilde{\chi}_4 \quad\mbox{and}\quad \tilde{\partial Q}^2\equiv\left(\frac{\partial \tilde{Q}\left(x\right)}{\partial \log{x}}\right)^2,
\end{equation}
\noindent
and the lag dependence in the amplitudes 
\begin{equation}
h_{\phi}\left(\tau\right) \quad\mbox{and}\quad \partial\zeta_{\phi}^2\equiv\left(\frac{\partial\log{\zeta_{\phi}\left(\tau\right)}}{\partial\log{\rho}}\right)^2.
\end{equation}
\noindent
These two sets of functions can then be compared independently. The shape terms represented on top of Fig.~(\ref{Figshape}) were obtained by averaging over time. They appear to match very neatly, and the overall factor is close to $\tilde{\chi_4}\approx 5 \tilde{\partial Q}^2$. This factor is the squared maximum slope of $\tilde{Q}$ and would be equal to $4$ if the statistics of the displacements, in addition to obey the above scaling, were also gaussian. We can now turn to the comparison of the lag-dependent amplitudes $h_{\phi}$ and $\partial\zeta_{\phi}^2$ represented on the middle panel of Fig.~(\ref{Figshape}) for three values of the packing fraction. These amplitudes have been normalized by their maximum for proper comparison, and extend over the range of lags where derivation can be accurately achieved. Surprisingly, we observe that the response term $\partial\zeta_{\phi}^2$ is non-monotonic and that these amplitudes reach their maximum for very similar lags. As stated above, the existence of the maxima in the response term is directly related to the existence of the superdiffusive regime: as readily visible from the sketch of Fig. 1 (Top), it confers a characteristic bell shape to $d\sigma_{\phi}(t)/d\phi \propto d\zeta_{\phi}(t)/d\phi$.

\begin{figure}[t]
\centering
\includegraphics[width=\columnwidth]{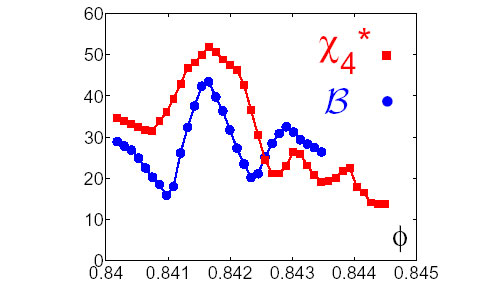}
\caption{Final comparison of the maximum dynamical susceptibility $\chi_{4}^*\left(\blacksquare\right)$ with the bound $\mathcal{B}\left(\newmoon\right)$ for the studied packing fractions.}
\label{fig:FigSoleBorne}
\vspace{0mm}
\end{figure}

The last step in investigating the quality of (\ref{ine}), is to compare the absolute $\phi$-dependent bound
\begin{equation}
\mathcal{B}\left(\phi\right)\equiv\frac{N\langle \delta \rho^2 \rangle}{\left\langle \rho\right\rangle^2} \mbox{max}_{\tau}\partial\zeta_{\phi}^2~~\mbox{max}_{x}\tilde{\partial Q}^2,
\label{bb}
\end{equation}  
\noindent
and the maximum dynamical susceptibility (recall that $\max_x\tilde{\chi}_4(x)=1$):
\begin{equation}
\chi_4^*\left(\phi\right)\equiv\mbox{max}_{\tau}h_{\phi}\left(\tau\right).
\label{cc}
\end{equation}
\noindent
We have measured the density fluctuations $\langle \delta \rho^2 \rangle$ within circular regions of increasing radius $R$ and asymptotically found the expected behavior $\langle \delta \rho^2 \rangle\left(R\right)\propto R^{-2}\propto N^{-1}$ as can be seen on the insert of the bottom left panel of Fig.~(\ref{Figshape}). As expected from a naive compressibility argument, the corresponding intensive normalized amplitude ${N\langle \delta \rho^2 \rangle}/{\left\langle \rho\right\rangle^2}$ is a decreasing function of the packing fraction. The final evaluation of the terms (\ref{bb}) and (\ref{cc}) to be compared is shown on Fig.~(\ref{fig:FigSoleBorne}) for packing fractions allowing proper evaluation. The lower bound $\mathcal{B}$ is observed to match the dynamical susceptibility quite accurately, in particular it reproduces remarkably well the peak of dynamical susceptibility across the jamming transition.  The fact that it is found to be larger than $\chi_4$ for high densities is probably due to a loss of accuracy in our numerical derivatives and in the evaluation of the density fluctuations in that regime.

\section{Discussion}
Let us now discuss these results physically. First, we have uncovered a strong scaling property in the wavelength dependence of both the dynamical susceptibility and response, which emphasizes the fact that dynamical fluctuations at a certain time scale are maximum when probed at the corresponding diffusive length scale. However, the remaining prefactor $h_{\phi}\left(\tau\right)$ still exhibits a maximum as a function of $\tau$. Therefore, evaluating  this quantity at {\it fixed} probe-length or wave-vector does not in itself lead to the identification of the dynamic fluctuation time. Second, we have found that the temporal behavior of the bound, driven by the derivative of the diffusion length with respect to the packing fraction, is a non-monotonic function of time. This behavior is due to  superdiffusion, as sketched in Fig.~\ref{figone}. The fact that both sides of the inequality reach their temporal maximum at comparable lags establishes a direct link between dynamical heterogeneity and anomalous autocorrelation of the grains displacements that leads to superdiffusion. The moving regions then appear to be disordered spatially but with persistent currents with typical life time $\tau_{sD}$. Finally, the lower bound was found to account for a substantial fraction of the equality (\ref{eqa}). This gives support to the hypothesis we made to describe the statistics of the system, i.e. the absence of long-range correlations in the density field, and suggests the existence of a large deviation function for the intensive observables. Perhaps more importantly, the quality of the bound shows that density fluctuations is the main determinant of the statistical properties of the dynamics in our system, which indicates that the intensity of energy injection plays a minor role in this range of packing fractions. Finally, the bound is found to be a very accurate indicator for the maximum amplitude reached by the dynamical susceptibility $\chi_4$ at each density.

Altogether the present work provides with the first experimental investigation of a dynamical fluctuation-response relation which gives a lower bound on the amplitude of dynamical heterogeneities. The bound reproduces the dependence of the dynamical susceptibility on probe-length, lag time, and correctly captures its density variation and amplitude. These important results clearly establish the relevance of this bound for characterizing dynamical heterogeneity in an athermal system. It would now be interesting to clarify the respective contributions of temperature and density in the context of thermal glasses. The next step would be to extract the ``fixed density'' contribution to $\chi_4$ in Eq. (\ref{eqa}) and understand in detail its physical origin.

\end{document}